\documentclass[prl,twocolumn,superscriptaddress]{revtex4}
\usepackage{amssymb}
\usepackage{makeidx}
\usepackage{amsmath}
\usepackage{graphicx}
\usepackage{dcolumn}
\usepackage{bm}
\usepackage{epstopdf}
\usepackage{color}
\usepackage{txfonts}
\usepackage{microtype}

\begin{document}

\title{High harmonic generation in fullerene molecules}
\author{H. K. Avetissian}
\author{A. G. Ghazaryan}
\author{G. F. Mkrtchian}
\email{mkrtchian@ysu.am}
\affiliation{Centre of Strong Fields Physics, Yerevan State University, 0025, Yerevan,
Armenia}

\begin{abstract}
Using dynamical Hartree-Fock mean-field theory, we study the high-harmonic
generation (HHG) in the fullerene molecules C$_{60}$ and C$_{70}$ under
strong pump wave driving. We consider a strong-field regime and show that
the output harmonic radiation exhibits multiple plateaus, whose borders are
defined by the molecular excitonic lines and cutoff energies within each
plateau scale linearly with the field strength amplitude. In contrast to
atomic cases for the fullerene molecule, with the increase of the pump wave
photon energy the cutoff harmonic energy is increased. We also show that
with the increase of the electron-electron interaction energy overall the
HHG rate is suppressed. We demonstrate that the C$_{70}$ molecule shows
richer HHG spectra and a stronger high-harmonic intensity than the C$_{60}$.
\end{abstract}

\date{\today }
\maketitle

\textit{Introduction}-- An intense light interaction with a quantum system
can excite the system's electrons towards extreme nonequilibrium states
during a fraction of its cycle \cite{Avetissian,Piazza}. Excited by the
wavefield and subjected to the internal forces inside the system, the
electrons emit coherent electromagnetic (EM) radiation that can contain from
tens to many hundreds of harmonics of an incident light \cite%
{Agostini2004,Kohler2012}. This is one of the fundamental processes in the
intense laser-matter interaction called high harmonic generation (HHG) \cite%
{Corkum}. The HHG process in atoms or molecules with the three-step model 
\cite{three-step} explanation is a well-demonstrated method for producing
coherent extreme ultraviolet radiation. The coherent spectrum of HHG implies
access to the extreme time resolution of the underlying quantum dynamics
that opens the way for attosecond physics \cite{Atto1,Atto2} and ultrafast
imaging methods for emitters themselves. In particular, using HHG
spectroscopy one can reconstruct the crystal potential \cite{Lakhotia},
observe Mott \cite{Mott} and Peierls \cite{Peierls} transitions, retrieve
the band structure \cite{Vampa,Dejean}.

For HHG it is crucial to increase HHG conversion efficiency and to extend
the harmonics cutoff \cite{Wah}. Specifically, the conversion efficiency of
the HHG process strongly depends on the density of emitters and the density
of states of emitters. The use of molecular systems, clusters, and crystals
can significantly increase the harmonic intensity by utilizing multiple
excitation channels \cite{Donnelly96,Vozzi05,Smirnova09}. Thus, in the last
decade, there has been a growing interest to extend HHG to crystals \cite%
{sol1,Zaks,Langer,sol2,sol3,sol4,sol5,sol6,sol7} and two-dimensional
nanostructures, such as semimetallic graphene \cite%
{gr1,gr2,gr3,gr4,gr5,gr6,gr7,gr8,gr9,gr10,gr11,gr12}, semiconductor
transition metal dichalcogenides \cite{TMD1,TMD2}, and dielectric hexagonal
boron nitride \cite{BN}. Currently, this is a new growing research field --
extreme nonlinear optics of nanostructured materials.

Among the variety of nanostructured materials, carbon allotropes play a
central role. The discovery of fullerene C$_{60}$ \cite{c60} through laser
evaporation of graphite and its synthesis in macroscopic amounts \cite{c60mc}
was triggered the study of many other carbon nanostructures, such as carbon
nanotubes \cite{nanotub}, graphene and its derivatives \cite{derivative}.
Currently, carbon nanomaterials are promising materials for many
applications, and in particular for extreme nonlinear optics. Being the
member of the carbon allotropes, it is expected a strong HHG from fullerene
molecules. Experimentally, in Refs. \cite{Ganeev1,Ganeev2} it is reported a
strong harmonic signal from C$_{60}$ plasma. Theoretical works predicted a
strong HHG from a C$_{60}$ molecule \cite{Zhang1,Zhang2,Zhang3} and solid C$%
_{60}$ \cite{Zhang4}. The theoretical analyses in Refs. \cite%
{Zhang1,Zhang2,Zhang3} are dominated by a single-particle picture, but it is
unclear how the electron-electron Coulomb interaction leaves its mark on HHG
and sub-cycle electronic response in these materials. Besides, in \cite%
{Zhang1}, moderately strong fields were considered, and excitonic lines were
termed as noninteger harmonics. However, these intrinsic exciton lines are
the result of Raman scattering of light, not harmonic radiation. Another
problem is how the symmetry groups of the most abundant fullerenes C$_{60}$
and C$_{70}$, namely the icosahedron group and the dihedral group, affect
the HHG process in these materials.

In the present work, we develop a microscopic theory of a fullerene molecule
nonlinear interaction with strong EM radiation of linear polarization taking
into account electron-electron interaction (EEI). In particular, we consider
C$_{60}$ and C$_{70}$ molecules as the most abundant examples of fullerene
molecules with different point group symmetries. By means of the dynamical
Hartree-Fock approximation, we reveal the general and basal structure of the
HHG spectrum and its relation to the molecular excitations. 
\begin{figure*}[tbp]
\includegraphics[width=.90\textwidth]{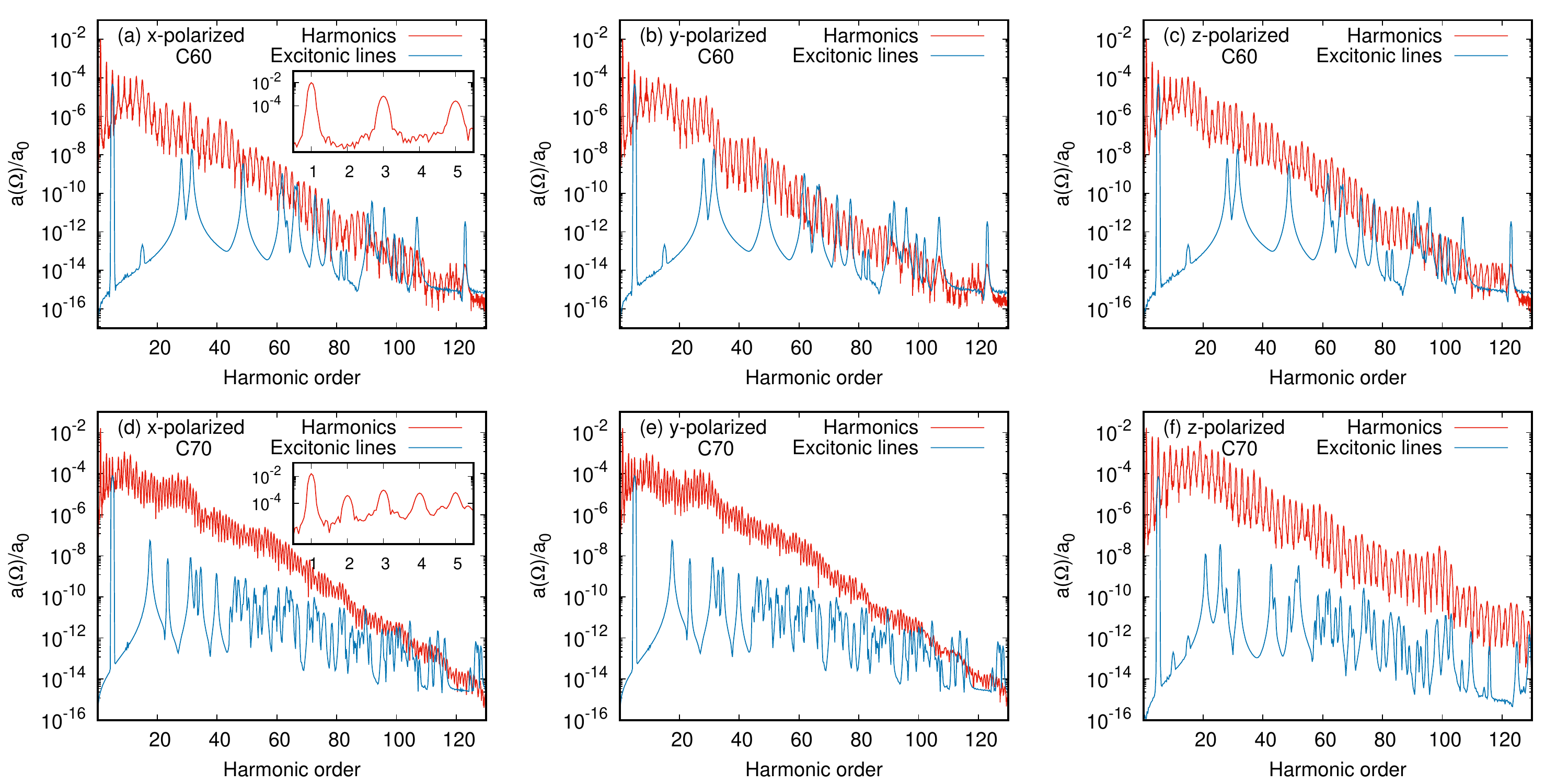}
\caption{The HHG spectra in the strong-field regime in logarithmic scale via
the normalized dipole acceleration Fourier transformation $a\left( \Omega
\right) /a_{0}$ (in arbitrary units) for C$_{60}$ (a, b, c) and for C$_{70}$
(d, e, f). The wave is assumed to be linearly polarized along the different
axes. The frequency is $\protect\omega =0.1\ \mathrm{eV}/\hbar $ and the
field strength is taken to be $E_{0}=0.5\ \mathrm{V/\mathring{A}.}$The
relaxation rate is taken to be $\hbar \protect\gamma =50\ \mathrm{meV}$. It
is also shown the molecular excitonic lines (lower curves, except the line
near $0.5\ \mathrm{eV}$). The latter is obtaind at the excitation of C$_{60}$
and C$_{70}$ with probe laser pule of frequency $0.5\ \mathrm{eV}/\hbar $
and $E_{0}=10^{-4}\ \mathrm{V/\mathring{A}}$\textrm{. }The relaxation rate
is taken to be $\hbar \protect\gamma =0.5\ \mathrm{meV}$. The spectra are
shown for moderate EEI energy: $U=2\ \mathrm{eV}$.}
\end{figure*}

\textit{Model}-- Let a fullerene molecule, C$_{60}$\ or C$_{70}$, interact
with strong coherent EM radiation that results in HHG. We assume neutral
fullerene molecules, which will be described in the scope of the
tight-binding theory where the interball hopping is much smaller than the
on-ball hopping, and EEI is described in the extended Hubbard approximation 
\cite{HV1,HV2,HV3,HV4}. Hence, the total Hamiltonian reads:\textrm{\ }%
\begin{equation}
\widehat{H}=\widehat{H}_{0}+\widehat{H}_{\mathrm{int}},  \label{H1}
\end{equation}%
where%
\begin{equation}
\widehat{H}_{0}=-\sum_{\left\langle i,j\right\rangle \sigma
}t_{ij}c_{i\sigma }^{\dagger }c_{j\sigma }+\frac{U}{2}\sum_{i\sigma
}n_{i\sigma }n_{i\overline{\sigma }}+\frac{1}{2}\sum_{\left\langle
i,j\right\rangle }V_{ij}n_{i}n_{j}  \label{Hfree}
\end{equation}%
is the free fullerene Hamiltonian. Here $c_{i\sigma }^{\dagger }$\ creates
an electron with the spin polarization $\sigma =\left\{ \uparrow ,\downarrow
\right\} $\ at site $i$\ ($\overline{\sigma }$\ is the opposite to $\sigma $%
\ spin polarization), and $\left\langle i,j\right\rangle $\ runs over all
the first nearest-neighbor hopping sites with the transfer energy $t_{ij}$.
The density operator is: $n_{i\sigma }=c_{i\sigma }^{\dagger }c_{i\sigma }$,
and the total electron density for site $i$\ is: $n_{i}=n_{i\uparrow
}+n_{i\downarrow }$. The second and third terms in (\ref{Hfree}) describe
the EEI Hamiltonian with on-site and inter-site Coulomb repulsion energies $U
$\ and $V_{ij}$, respectively. Since the distance $d_{ij}$\ between the
nearest-neighbour pairs varies over the system, we scale inter-site Coulomb
repulsion: $V_{ij}=Vd_{\min }/d_{ij}$, where $d_{\min }$\ is the minimal
nearest-neighbor distance. For the all calculations we use the ratio $V=0.4U$%
\ \cite{HV1,HV3}. In the Hamiltonian, we neglected the lattice vibrations.
In the calculations, the light-matter interaction is described in the
length-gauge via the pure scalar potential%
\begin{equation}
\widehat{H}_{\mathrm{int}}=e\sum_{i\sigma }\mathbf{r}_{i}\cdot \mathbf{E}%
\left( t\right) c_{i\sigma }^{\dagger }c_{i\sigma },  \label{3}
\end{equation}%
with the elementary charge $e$, position vector $\mathbf{r}_{i}$, and the
electric field strength $\mathbf{E}\left( t\right) =f\left( t\right) E_{0}%
\hat{\mathbf{e}}\cos \omega t$, with the frequency $\omega $, polarization $%
\hat{\mathbf{e}}$ unit vector, and pulse envelope $f\left( t\right) =\sin
^{2}\left( \pi t/\mathcal{T}\right) $. The pulse duration $\mathcal{T}$ is
taken to be $20$ wave cycles: $\mathcal{T}=40\pi /\omega $. From the
Heisenberg equation one can obtain evolutionary equations for the
single-particle density matrix $\rho _{ij}^{\left( \sigma \right)
}=\left\langle c_{j\sigma }^{\dagger }c_{i\sigma }\right\rangle $. In
addition, we will assume that the system relaxes at a rate $\gamma $ to the
equilibrium $\rho _{0ij}^{\left( \sigma \right) }$ distribution. To obtain a
closed set of equations for the single-particle density matrix $\rho
_{ij}^{\left( \sigma \right) }=\left\langle c_{j\sigma }^{\dagger
}c_{i\sigma }\right\rangle $, EEI will be considered under the Hartree-Fock
approximation \cite{Supplement}. 
\begin{figure*}[tbp]
\includegraphics[width=.90\textwidth]{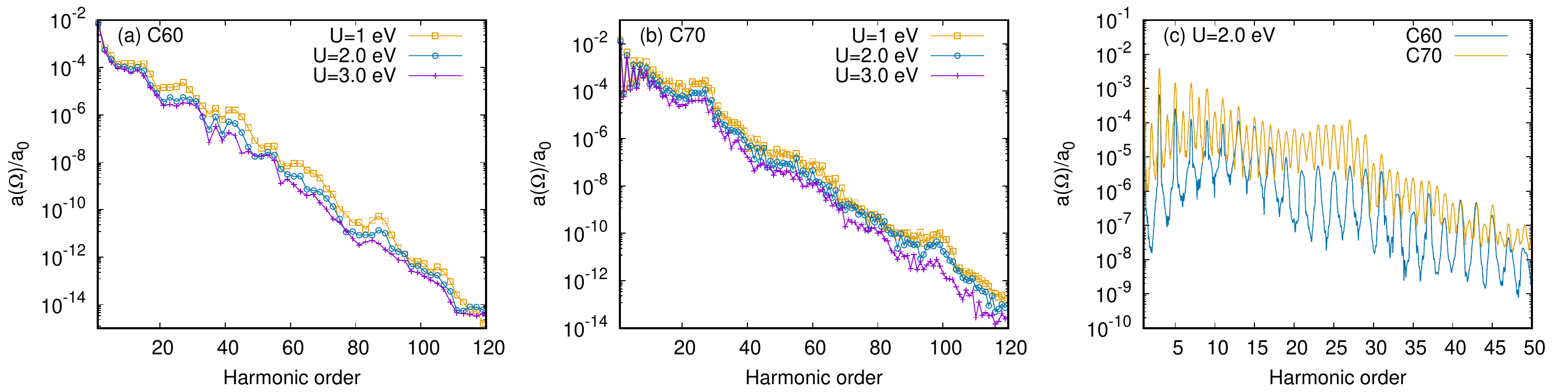}
\caption{The envelopes of HHG spectra (only peaks are connected) in the
strong-field regime for different EEI energies in logarithmic scale via the
normalized dipole acceleration Fourier transformation $a\left( \Omega
\right) /a_{0}$ (in arbitrary units) for C$_{60}$ (a) and C$_{70}$ (b). The
frequency is $\protect\omega =0.1\ \mathrm{eV}/\hbar $ and the field
strength is taken to be $E_{0}=0.5\ \mathrm{V/\mathring{A}}$\textrm{. }The
relaxation rate is taken to be $\hbar \protect\gamma =50\ \mathrm{meV}$. In
(c) we show HHG for C$_{60}$ versus C$_{70}$. }
\end{figure*}
Thus, we obtain the following equation for the density matrix: 
\begin{equation*}
i\hbar \frac{\partial \rho _{ij}^{\left( \sigma \right) }}{\partial t}%
=\sum_{k}\left( \tau _{kj\sigma }\rho _{ik}^{\left( \sigma \right) }-\tau
_{ik\sigma }\rho _{kj}^{\left( \sigma \right) }\right) +\left( V_{i\sigma
}-V_{j\sigma }\right) \rho _{ij}^{\left( \sigma \right) }
\end{equation*}

\begin{equation}
+e\mathbf{E}\left( t\right) \left( \mathbf{r}_{i}-\mathbf{r}_{j}\right) \rho
_{ij}^{\left( \sigma \right) }-i\hbar \gamma \left( \rho _{ij}^{\left(
\sigma \right) }-\rho _{0ij}^{\left( \sigma \right) }\right) ,  \label{evEqs}
\end{equation}%
where $V_{i\sigma }=\sum_{j\alpha }V_{ij}\left( \rho _{jj}^{\left( \alpha
\right) }-\rho _{0jj}^{\left( \alpha \right) }\right) +U\left( \rho
_{ii}^{\left( \overline{\sigma }\right) }-\rho _{0ii}^{\left( \overline{%
\sigma }\right) }\right) $\ and $\tau _{ij\sigma }=t_{ij}+V_{ij}\left( \rho
_{ji}^{\left( \sigma \right) }-\rho _{0ji}^{\left( \sigma \right) }\right) $%
\ are defined via$\ $the$\ $density matrix $\rho _{ij}^{\left( \sigma
\right) }$\ and its initial value.

We numerically diagonalize the tight-binding Hamiltonian $\widehat{H}_{0}$
with the parameters that provide molecular orbitals close to experiment \cite%
{H3,gap1,Supplement}, and construct the initial density matrix $\rho
_{0ij}^{\left( \sigma \right) }$ via the filling of electron states in the
valence band according to the zero temperature Fermi--Dirac-distribution $%
\rho _{0ij}^{\left( \sigma \right) }=\sum_{\mu =N/2}^{N-1}\psi _{\mu }^{\ast
}\left( j\right) \psi _{\mu }\left( i\right) $, where $\psi _{\mu }\left(
i\right) $ is the eigenstate of $\widehat{H}_{0}$.

\textit{Results-- }The HHG spectrum is evaluated from the Fourier
transformation $\mathbf{a}\left( \Omega \right) $ of the dipole acceleration 
$\mathbf{a}\left( t\right) =d^{2}\mathbf{d}/dt^{2}$. The dipole is defined
as $\mathbf{d}\left( t\right) =e\sum_{i\sigma }\mathbf{r}_{i}\rho
_{ii}^{\left( \sigma \right) }\left( t\right) $. For convenience, we
normalize the dipole acceleration by the factor $a_{0}=e\overline{\omega }%
^{2}\overline{d},$ where $\overline{\omega }=1\ \mathrm{eV}/\hbar $ and $%
\overline{d}=1\ \mathrm{\mathring{A}}$. The power radiated at the given
frequency is proportional to $\left\vert \mathbf{a}\left( \Omega \right)
\right\vert ^{2}$.

\begin{figure}[tbp]
\includegraphics[width=.44\textwidth]{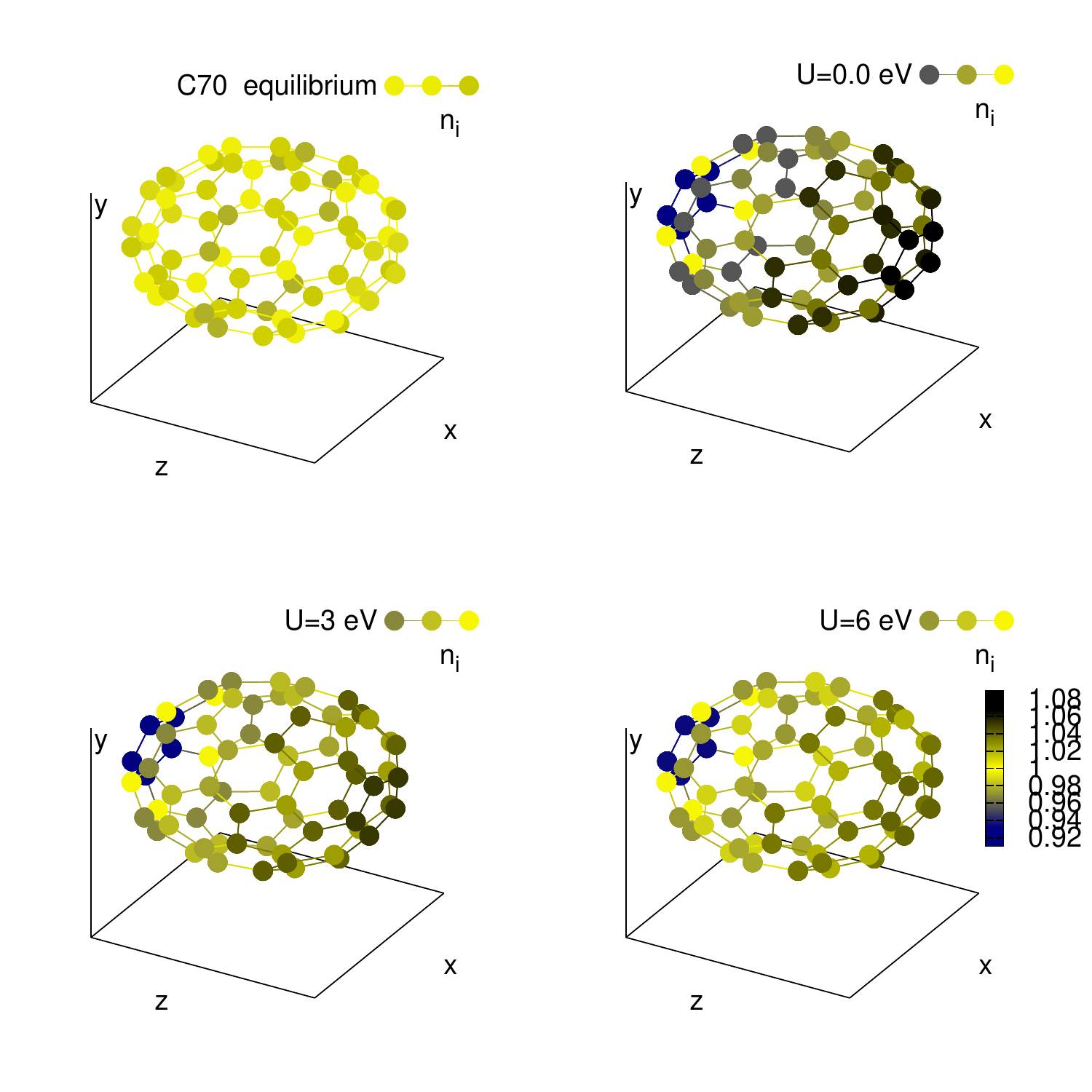}
\caption{The site occupations $n_{i}$ in 3D color mapped molecular
structures for C$_{70}$. The wave is assumed to be linearly polarized along
the z-axis. \ The first configuration is the equilibrium, the next 3
configurations are at the peak of the laser field $t=\mathcal{T}/2$ for
different EEI energies.}
\end{figure}

In order to clarify main aspects of HHG in C$_{60}$ and C$_{70}$, we assume
the excitation frequency is much smaller than the typical gap $\sim 2\ eV$.
For all calculations the wave is assumed to be linearly polarized. Orienting
the linearly polarized pump wave along different axes results in different
harmonics spectra. The difference is essential for C$_{70}$. This is because
for C$_{70}$ the inversion symmetry takes place only for the 5-fold rotation
axes (z-axis). In Fig. 1(a-f), we show the HHG spectra in the strong field
regime ($E_{0}=0.5\ V/\mathring{A}$) for different orientation of the wave's
polarization. For C$_{60}$\ molecule, because of the inversion symmetry only
the odd harmonics appear in the HHG spectrum. For C$_{70}$\ molecule we have
both even and odd harmonics except at the z-polarization of the pump wave.
The insets in Figs. 1(a) and 1(d) show low harmonics, where the difference
in symmetry for C$_{60}$\ and C$_{70}$\ is clearly visible (see, also \cite%
{Supplement} for a relatively weak field). As is seen from Fig. 1, in both
cases HHG spectra have a multi plateau structure that is connected with the
intrinsic molecular excitations between the occupied molecular orbitals and
the unoccupied molecular orbital. To show this in a more transparent way, in
Fig. 1 it is also shown the molecular excitonic lines. The latter is
obtained at the excitation of C$_{60}$\ and C$_{70}$\ with probe weak laser
pulse. In particular, the lines near 2.7 $eV$\ and 1.7 $eV$\ for C$_{60}$\
and C$_{70}$, respectively, are the first dipole-allowed transition from
highest occupied molecular orbital to the lowest unoccupied molecular
orbital. These excitonic lines were termed as noninteger harmonics in \cite%
{Zhang1} have their fingerprints in the multi plateau structure for the HHG
spectra in the strong-field regime. As is seen, plateaus' borders are
defined by these lines. Besides for C$_{70}$\ there are close lines which
enhance the HHG yield compared with C$_{60}$.

Since the boundaries of the plateaus are determined by dipole-allowed
multiple electronic transitions between the molecular orbitals, it can be
argued that for the frequencies $\hbar \omega <<$\ $t_{ij}$\ these
boundaries are almost invariant with respect to the pump wave frequency.
They are determined through the intrinsic features of the free fullerene.
Note also that the position of excitonic lines and relative intensities
depend also on EEI. Besides, as was shown in \cite{Zhang5}, the on-site EEI
suppresses the charge fluctuation and reduces the absorbed energy. It is
also expected HHG yield suppression due to EEI. The latter is shown in Fig.
2, where the HHG spectra in the strong-field regime for different EEI
energies are shown. To obtain the mean picture which does not depend on the
orientation of the molecule with respect to laser polarization we take the
wave polarization unit vector as $\hat{\mathbf{e}}=1/\sqrt{3}\left\{
1,1,1\right\} $. To make the plateaus more visible in Figs. 2(a) and 2(b) we
show the envelopes of HHG spectra. As is seen from Figs. 2(a) and 2(b), with
the increase of the EEI energy overall the HHG rate is suppressed. Another
interesting aspect of HHG in fullerene is the qualitative and quantitative
difference between both molecules. As is seen, the C$_{70}$ molecule shows
more pronounced nonlinear properties (Fig. 2(c)). Due to broken inversion
symmetry in the case of C$_{70}$, we have even harmonics, besides due to the
smaller energy gap and the larger density of states the HHG rate is larger
by one to two orders compared with C$_{60}$. For comparison, in Fig. 2(c) we
show the first four plateaus for both molecules. The suppression of HHG with
the increase of EEI energy is connected with the fact that at strong on-site
and inter-site electron-electron repulsion the polarizability of molecules
or in other words electrons migration from the equilibrium states is
suppressed. To show this visually, in Fig. 3 we display site occupations $%
n_{i}=\left\langle c_{i\uparrow }^{\dagger }c_{i\uparrow }\right\rangle
+\left\langle c_{i\downarrow }^{\dagger }c_{i\downarrow }\right\rangle $ via
3D color mapped molecular structures at the peak of the laser field for
z-polarized wave for the same interaction parameters. The color bar
represents site occupation. As can be seen from Fig. 3, the deviation from
the equilibrium position and, therefore, the polarization is maximum for
vanishing EEI. 
\begin{figure*}[tbp]
\includegraphics[width=.92\textwidth]{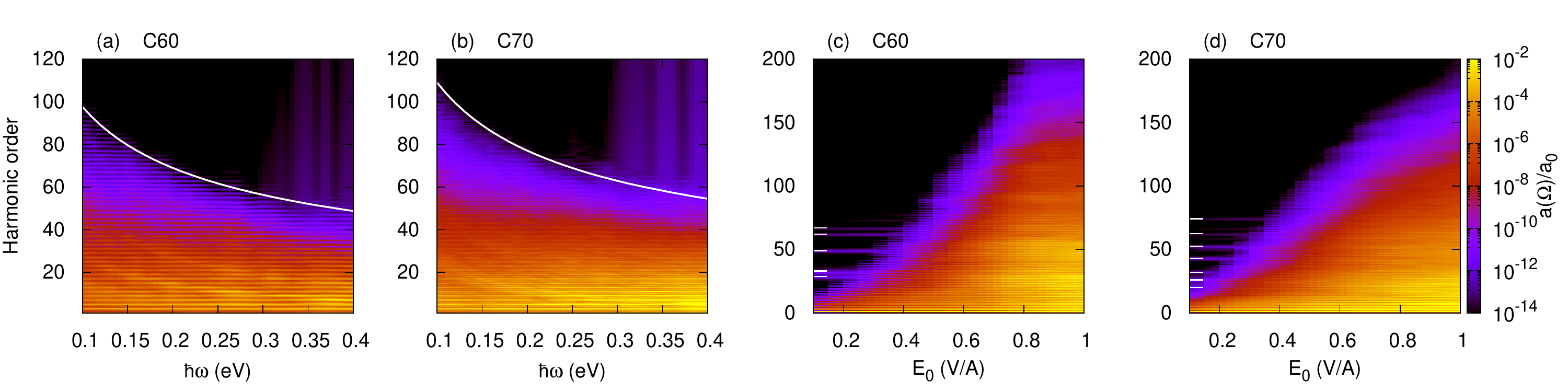}
\caption{The HHG spectra versus pump wave frequency (a,b) and intensity
(c,d). The color bar represents the emission rate via dipole acceleration
Fourier transformation $a\left( \Omega \right) /a_{0}$ in the logarithmic
scale for C$_{60}$ (a, c) and C$_{70}$ (b,d). The wave is assumed to be
linearly polarized with polarization unit vector $\hat{\mathbf{e}}=\frac{1}{%
\protect\sqrt{3}}\left\{ 1,1,1\right\} $. The field strength is fixed $%
E_{0}=0.5\ \mathrm{V/\mathring{A}}$ for (a) and (b), while for (c) and (d)
the frequency is fixed $\protect\omega =0.1\ \mathrm{eV}/\hbar \mathrm{.}$%
The spectra are shown for moderate EEI energy: $U=2\ \mathrm{eV}$. The
relaxation rate is taken to be $\hbar \protect\gamma =50\ \mathrm{meV}$. The
white lines in (a) and (b) are envelopes $\protect\alpha /\protect\sqrt{%
\protect\omega }$ which show the cutoff harmonic positions. The white
horizontal lines in (c) and (d) are excitonic resonances that show each
plateau's borders.}
\end{figure*}
\begin{figure*}[tbp]
\includegraphics[width=.92\textwidth]{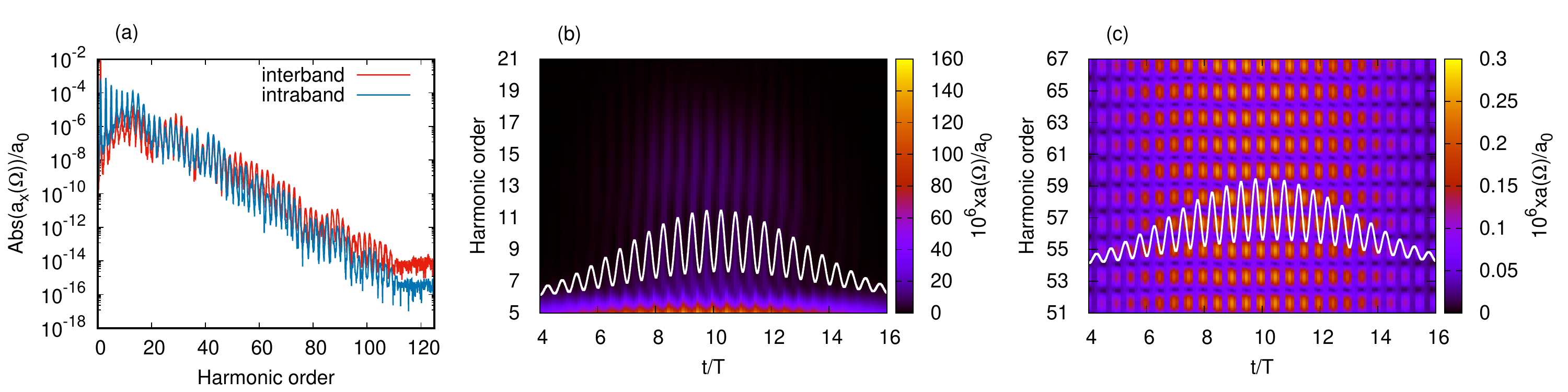}
\caption{The interband and itraband contribution in HHG spectra (a) and the
spectrogram (b,c) of the HHG process via the windowed Fourier transform of
the dipole acceleration for C$_{60}$. The wave is assumed to be linearly
polarized with polarization unit vector $\hat{\mathbf{e}}=\frac{1}{\protect%
\sqrt{3}}\left\{ 1,1,1\right\} $. The frequency is $\protect\omega =0.1\ 
\mathrm{eV}/\hbar $ and the field strength is taken to be $E_{0}=0.5\ 
\mathrm{V/\mathring{A}.}$ The spectra are shown for $U=2\ \mathrm{eV}$ and $%
\hbar \protect\gamma =50\ \mathrm{meV}$. (b) low frequency part and (c) high
frequency part. Then, the white curve on the density plots is scaled by the
factor 10 population of the initially unoccupied molecular orbitals versus
time.}
\end{figure*}

We also investigated the HHG spectra dependence versus pump wave frequency
and intensity. In Figs. 4(a) and 4(b) we plot the HHG spectra versus pump
wave frequency at moderate EEI energy $U=2$ $\mathrm{eV}$, for C$_{60}$ and C%
$_{70}$, respectively. For both molecules, the cutoff harmonic position is
well approximated by the dependence $N_{\mathrm{cut}}\sim \omega ^{-1/2}$
which is plotted along with density plot. Note that for atomic HHG via free
continuum $N_{\mathrm{cut}}\sim \omega ^{-3}$ \cite{three-step}. In case of
two-level atom $N_{\mathrm{cut}}\sim \omega ^{-1}$ \cite{2Lhhg}. Thus, in
contrast to atomic cases for fullerene molecule with the increase of the
pump wave photon energy the cutoff harmonic energy ($\hbar \omega N_{\mathrm{%
cut}}$) is increased.

Next, we consider the HHG spectra as a function of pump wave intensity. In
Figs. 4(c) and 4(d), we show the HHG spectra as a function of field
amplitude and the harmonic order for a fixed frequency. The HHG spectra have
interesting structures. First of all, it is clearly seen excitonic lines
(marked by the white horizontal lines). These excitonic lines define the
borders of the plateaus and within each plateau, the cutoff harmonic
linearly increases with increasing the field strength. Then, reaching the
harmonic $\sim $160 which corresponds to the transition of the lowest
occupied molecular orbital to the highest unoccupied molecular orbital, the
HHG rate is saturated. Note that linear dependence of the cutoff harmonics
on the field strength is inherent to HHG via discrete levels \cite{2Lhhg},
or in crystals with linear energy dispersion \cite{sol3,sol4,gr11}.

We now consider the origin of the HHG in fullerene molecules. There are two
contributions to the current: the electron/hole transitions within
unoccupied/occupied molecular orbitals and the electron-hole creation
(transitions from occupied molecular orbitals to unoccupied ones) and
subsequent recombination. The former makes contribution only for low
harmonics and is analogous to the intraband current in a semiconductor,
while the latter makes the main contribution in the high frequency part
corresponds to the interband current, which represents
recombination/creation of electron-hole pairs. This picture is analogous to
HHG in solid state systems. To separate these contributions in the dipole
acceleration spectrum we made change of the basis via formula $\rho
_{ij}=\sum_{\mu ^{\prime }}\sum_{\mu }\psi _{\mu ^{\prime }}^{\ast }\left(
j\right) \rho _{\mu \mu ^{\prime }}\psi _{\mu }\left( i\right) $, where $%
\rho _{\mu \mu ^{\prime }}$ is the density matrix in the energetic
representation. Hence, we define interband part of dipole acceleration, as 
\begin{equation}
\mathbf{d}_{\mathrm{inter}}\left( t\right) =2\sum_{\mu ^{\prime
}=N/2}^{N-1}\sum_{\mu =0}^{N/2-1}\mathrm{Re}\left( \rho _{\mu \mu ^{\prime
}}\left( t\right) \mathbf{d}_{\mu ^{\prime }\mu }\right) ,  \label{dinter}
\end{equation}%
and intraband part will be 
\begin{equation}
\mathbf{d}_{\mathrm{intra}}\left( t\right) =\sum_{\mu ,\mu ^{\prime
}=N/2}^{N-1}\rho _{\mu \mu ^{\prime }}\left( t\right) \mathbf{d}_{\mu
^{\prime }\mu }+\sum_{\mu ,\mu =0}^{N/2-1}\rho _{\mu \mu ^{\prime }}\left(
t\right) \mathbf{d}_{\mu ^{\prime }\mu },  \label{dintra}
\end{equation}%
where dipole transition matrix elements are $\mathbf{d}_{\mu ^{\prime }\mu
}=e\sum_{i}\psi _{\mu ^{\prime }}^{\ast }\left( i\right) \mathbf{r}_{i}\psi
_{\mu }\left( i\right) $. In Fig. 5(a), we show the interband/itraband
contribution in HHG spectra for C$_{60}$. The similar picture we have for C$%
_{70}$. As is seen, intraband dipole acceleration is significant for low
frequency part of the spectrum, while in the high frequency part the main
contribution is caused by the electron-hole creation and subsequent
recombination. This information can also be extracted from the evolution of
the high harmonic spectrum as a function of time. For this propose a
Blackman window of width $1.2\pi /\omega $ is scanned across $20$ optical
cycles. The results along with the population of the conduction band
(unoccupied molecular orbitals) $W\left( t\right) =\sum_{\mu =0}^{N/2-1}\rho
_{\mu \mu }\left( t\right) $ are displayed in Figs. 5(b) and 5(c). As is
seen from these figures, the emission of high harmonics takes place two
times per wave cycle, corresponding to two maxima or minima of the
population. The low frequency harmonic bursts take place in-between maxima
and minima of the population Fig. 5(b), while higher harmonics are the
results of the recombination and the bursts take place at minima of the
population (Fig. 5(c)). There are also a domain of harmonics where we have
interplay between intra- and interband emission.

\textit{Conclusion}-- We revealed the general features of the HHG in
fullerene molecule under strong-field driving. The HHG spectra show multiple
plateaus, which is explained by the recombination of electrons and holes
from molecular orbitals. Those are intrinsic molecular excitations between
the unoccupied molecular orbitals and the occupied molecular orbital. These
intrinsic molecular excitations -- so-called excitonic lines define the
borders of the plateaus. Within the each plateau, the cutoff harmonic
linearly increases with increasing the field strength. In contrast to atomic
cases, for fullerene molecule with the increase of the pump wave photon
energy the cutoff harmonic energy is increased. The HHG spectra strongly
depends on the molecule symmetry qualitatively as well as quantitatively.
The C$_{70}$ molecule shows more pronounced nonlinear properties due to
degradation of molecular symmetry compared with C$_{60}$. We also revealed
the role of EEI. With the increase of the EEI energy overall the HHG rate is
suppressed. The fullerene molecules are known to have different isomers with
other point-group symmetries. Therefore, they are interesting systems as a
new sources of HHG, and spectroscopy based on HHG might be useful to reveal
the symmetries and electron dynamics involved. Developing a detailed
understanding of the HHG in different classes of fullerene molecule is an
interesting topic for the future work.

This work was supported by the RA State Committee of Science in the frame of
the research project 20TTWS-1C010.

\end{document}